\documentclass[12pt]{article}
\renewcommand{\abstractname}{\ }
\begin{document}
\renewcommand{\abstractname}{\ }

\title{Diffusion Scattering of Waves is a Model of Subquantum Level?}

\author{E. M. Beniaminov}
\date{}

\maketitle
\begin{abstract}
In the paper, we discuss the studies of mathematical models of diffusion scattering of waves in the phase space,
and relation of these models with quantum mechanics. In the previous works it is shown that in these models
of classical scattering process of waves, the quantum mechanical description arises as the asymptotics
after a small time. In this respect, the proposed models can be considered as examples in which the quantum
descriptions arise as approximate ones for certain hypothetical reality.
The deviation between the proposed models and the quantum ones can arise, for example, for processes
with rapidly changing potential function. Under its action the diffusion scattering process of waves will
go out from the states described by quantum mechanics.

In the paper it is shown that the proposed models of diffusion scattering of waves possess the property of
gauge invariance. This implies that they are described similarly in all inertial coordinate systems, i.~e.,
they are invariant under the Galileo transformations.

We propose a program of further research.
\end{abstract}

\section{Introduction}
Usually description of quantum systems is constructed by using formal quantization procedures, based on the
classical description of the corresponding mechanical systems. The search for the sense of these procedures
attracted many physicists, starting with A.~Einstein.

The interest to this subject periodically decreased and revived again.
In this direction, one can mention the von Neumann theorem, proved during the period of formation of quantum mechanics,
on impossibility of description of quantum mechanics by introducing hidden parameters
\cite{newmann}. Nevertheless, in the 50s, in the works of D.~Bohm and L.~de Broglie  \cite{bohm},  one proposed
a model of quantum mechanics with hidden parameters, not satisfying some conditions of the von Neumann theorem
and possessing a strange property of far-action. In the 60-s, in the work of E.~Nelson \cite{nelson}, one
proposed a probabilistic approach given the name ``stochastic quantization''. The subject of foundation
of quantum mechanics excited many specialists
(for example, D.~I.~Blohintsev \cite{blohintsev}, V.~P.~Maslov \cite{maslov}, K.~Popper \cite{poper}, etc.),
and it was discussed in their publications.
There are works in which one makes a detailed analysis of the problem of introducing hidden parameters
into quantum mechanics. They include a widely known work of Bell \cite{bell}
on introducing hidden parameters and non-locality of quantum mechanics. An interesting analysis of this work
is given in \cite{kchrennikov}. The paper \cite{cabello} contains a huge list of literature on foundations
of quantum mechanics, and provides a classification of these works. The history of the discussion around
the subject ``foundation of quantum mechanics'' and attitude to this subject of ``traditional physicists''
are remarkably described in the book by K.~Popper \cite{poper}.

The small popularity of alternative approaches to foundation of quantum mechanics among the
working physicists is usually related to the fact that they did not yet give serious new results.
They also did not give more convenience in computations and heuristics. However, recently the alternative approaches
cause again an intent attention related to the problems and possibilities of quantum optics,
as well as the problem of construction of quantum computers.

In the present paper we discuss the research on construction of models of diffusion scattering of waves in phase space.
I have been studying this subject during last years \cite{ben1, ben2, ben3, ben4}. In these models the quantum
description of processes arises as an approximate one, asymptotical for large values of certain coefficients of the
model.

In the papers mentioned above one makes an attempt to construct a model of quantum observables on the base of
wave functions on the phase space. Note that in quantum mechanics, the wave function depends either only on
coordinates or only on momenta, while in the present approach one considers wave functions depending both on
coordinates and on momenta. This model is based also on the following observation.
In quantum mechanics, the phase of the wave function of a particle (the natural hidden parameter) changes in time
even for stationary states with very high velocity (if one takes into account the stationary energy).
 This velocity is such that a transfer of the particle
with even small (non-relativistic) velocities can cause considerable changes in the phase of wave function
because of the relativistic effect of more slow inner processes of a moving particle. Already taking into
account this effect leads to non-commutativity of the action of coordinate and momentum shifts on the wave function.
Note once again that in the proposed model one considers wave functions on the phase (and not configuration) space,
and one assumes that the particle is in a diffusion process causing random shifts of the wave both by coordinates
and by momenta. It is shown that the classical model of scattering of the wave, taking into account the
assumptions described above, yields to arising quantum effects.

In the further sections of the present paper we speak in more detail on the obtained results and directions of
further research.

\section {The results obtained earlier}
In the paper \cite{ben1}
one introduces some assumptions on the process of observation of quantum phenomena, including introduction of
hidden parameters, action of the group of motion in the region of hidden parameters, and averaging
observations due to small random (diffusion) motions of the observed object. By an observable we mean,
as in classical mechanics,  an arbitrary integrable function $f(x,p)$   on the phase space $(x, p)\in R^{2n},$
where $x$ is the coordinate,  $p$ is the momentum. If $\rho (x, p)$ is the density of probability distribution
of the position of a particle in the phase space, then the mathematical expectation (mean) $\bar f$ of
an observable $f$ is given by the standard formula:
$$\bar f=\int \limits_{R^{2n}}f(x,p)\rho (x, p) dx dp.$$
Below it is assumed that in experiments, not all distributions $\rho (x, p)$ are realized, but only those of
the form
$\rho (x, p)=| \tilde\varphi(x, p)|^2, $  where $\tilde\varphi(x, p)$ is a wave function averaged in the diffusion process,
given in the form of a complex valued function $\varphi(x, p)$ on the phase space. It is shown that the functions of
the form $\tilde\varphi(x, p)$,
form a linear subspace $\cal H$ of ``stationary'' (averaged) wave functions in the space of all
square integrable functions on $R^{2n}$. Since $| \tilde\varphi|^2= \tilde\varphi^* \tilde\varphi$, where $\varphi^*$
is the complex conjugate function to the function $\varphi$, the mean value of the observable $f$
on averaged densities of probability distribution yields the following quadratic form on the space
of $\tilde\varphi \in\cal H$:

$$\bar f=\int \limits_{R^{2n}}f(x,p)\rho (x, p) dx dp=\int \limits_{R^{2n}}f(x,p)\tilde\varphi^*(x, p)\ \tilde\varphi(x, p) dx dp=
 \langle \tilde\varphi, A_f \tilde\varphi \rangle,$$
where by $A_f $ we denote the linear operator on the space $\cal H$ giving this quadratic form.

The introduced operator $A_f$ is called the operator of the observable $f$. It is natural that the spectrum
of this linear operator corresponds to possible values of observations for the observable $f$ under the
assumptions made.

In the paper \cite{ben1} we have found the expression for this operator for any observable (function of
coordinates and momenta), depending on the ratio $a/b$ of diffusion coefficients, with respect to coordinates
$a$ and momenta $b$, of the averaging process
of wave functions.
It is shown that the usual linear operator of a quantum observable does not coincide with the one constructed in the
paper, but differs by smoothing of the potential energy function with respect to the normal distribution
with the normal deviation equal to $\hbar a/2b$, where $\hbar$ is the Planck constant.
Assuming that this difference yields the shift of the spectrum of the hydrogen atom
observed in the Lamb experiment, we give an estimate of the quantity $a/b$.

A big advantage of the considered approach is also the possibility to express, for each wave function of the system,
the corresponding density of probability distribution in the phase space $\rho (x, p)=| \tilde\varphi(x, p)|^2$.
For the first time this problem was solved by Wigner \cite{wigner}, but he has constructed
``quasidistributions'' on the phase space which can be negative and hence have no physical sense.
And here we have a probability density distribution which is the result of smoothing of Wigner's
``quasidistribution'' with respect to the normal distribution with the normal deviation equal to
$\hbar a/2b$. Smoothed Wigner's distributions were first considered by Husimi \cite{husimi}, but the sense
of the smoothing parameters was unclear.

At the end of the paper \cite{ben1} we have posed the problems of generalization of the results to the
relativistically invariant case, taking into account the spin of the particles, more general phase manifolds,
and description of dynamics of observable quantities.

The papers \cite{ben2, ben3, ben4} are devoted to solution of the latter problem.

In the papers \cite{ben2, ben3} (in \cite{ben2} the results of \cite{ben3} have been announced),
continuing the work \cite{ben1}, we consider the classical model of the diffusion process for
a wave (complex valued) function $\varphi (x, p, t)$ on the phase space $(x, p)\in R^{2n}$
at the moment of time $t$. It is assumed that the wave function
 $\varphi (x, p, t)$ at the moment $t$ satisfies the following differential equation:
\begin{equation}\label{eq_diff1}
\frac{\partial\varphi}{\partial{t}}=\sum_{k=1}^{n}
\biggl(
\frac{\partial H}{\partial x_k} \frac{\partial\varphi}{\partial p_k}-
\frac{\partial H}{\partial p_k} \frac{\partial\varphi}{\partial x_k}
\biggr)
-\frac{i}{\hbar}
\biggl(H-\sum_{k=1}^{n}\frac{\partial H}{\partial p_k}p_k\biggr)\varphi
+\Delta_{a,b}{\varphi},
\end{equation}
\begin{equation}\label{delta}
\mbox{where }\ \ \ \ \ \Delta_{a,b}{\varphi}=
a^2\sum_{k=1}^{n}\biggl(\frac{\partial}{\partial{x_k}}-
     \frac{ip_k}{\hbar}\biggr)^{2}\varphi
+b^2\sum_{k=1}^{n}\frac{\partial^2 }{\partial{p^2_k}}\varphi
     +\frac {abn}{\hbar}{\varphi},
\end{equation}
where
$H(x, p)$ is the Hamilton function;
$a^2$ and $b^2$ are the diffusion coefficients with respect to coordinates and momenta respectively,
and $\hbar$ is the Planck constant.

The analysis of this equation has shown (see \cite{ben3}, Theorems 4 and~5) that in this model the motion splits into
rapid and slow ones. After the rapid motion, at the time of order $\hbar/(ab)$, starting from an arbitrary
wave function on the phase space, the system goes to a function belonging to certain special
subspace of ``stationary'' wave functions for the diffusion process.  The elements of this subspace
are parameterized by wave functions depending only on coordinates.
The slow motion takes place already in this subspace and is described by the Schrodinger equation, in which
in the right hand side we have the operator coinciding with the usual quantum mechanics Hamilton operator
up to summands of order $a\hbar /b$.

Thus, already in these papers it is shown that the quantum mechanical description of processes can arise
as the approximate description of the classical diffusion of waves in the phase space.
For the model considered in the paper, this approximation arises when the Hamilton function has a small change
with a change of coordinates, momenta and time in intervals of length of order defined by the Planck
constant and diffusion intensities.

Assuming the heat reason of the diffusions, in the paper we estimate the diffusion coefficients and the
transition time $\hbar/(ab)$ from the classical description of the process
in which the Heisenberg indeterminacy principle in general does not hold, to the quantum description in which
the Heisenberg principle already holds.
The transition time has order $1/T \cdot 10^{-11} sec$,
where $T$ is the temperature of the medium.

Another interesting result of the paper \cite{ben2} is that the solution of equation (\ref{eq_diff1}) can be
represented as a path integral, but not with respect to the Feynman ``measure'' \cite{feynman},
whose sense is mathematically not very much clear, but with respect to the probability measure
(analogous to the Wiener measure) for the Brownian motion given by the Fok\-ker--Planck equation of the form
\begin{eqnarray}\label{eq_diffuz}
{ \frac{\partial f}{\partial{t}}=} \sum_{k=1}^n \biggl (\frac{\partial H}{\partial x_k} \frac{\partial f}{\partial p_k}-
\frac{\partial H}{\partial p_k} \frac{\partial f}{\partial x_k}
+a^2 \frac{\partial^2 f}{\partial x_k^2}+ b^2 \frac{\partial^2 f}{\partial p_k^2}\biggr).
\end{eqnarray}
Here $f(x,p,t)$ is the probability density of the position of the Brownian particle in the phase space
at the moment of time $t$.
In this case, the sense of the path integral can be better substantiated.

Generalization of equation (\ref{eq_diff1}) to the relativistic case meets some difficulties, because
of the presence of diffusion with respect to coordinates in this model.
Such diffusions assume unbounded velocity in the diffusion jumps. Hence the next step in our investigations was
construction of a model of scattering of waves in the phase space, in which the diffusion takes place only with respect
to momenta, because of the collision with particles of the medium in the heat equilibrium.

In the paper \cite{ben4}, instead of equation (\ref{eq_diffuz}) we consider the Kramers equation \cite{kramers},
\cite{vankampen} of the form
\begin{equation}\label{eq_kramers}
\frac{\partial f}{\partial{t}}=\sum_{j=1}^{n}
\biggl(
\frac{\partial V}{\partial x_j} \frac{\partial f}{\partial p_j}-
 \frac {p_j}{m} \frac{\partial f}{\partial x_j}
\biggr)
+\gamma \sum_{j=1}^{n}\frac{\partial }{\partial{p_j}}\biggl (p_j f
+kTm\frac{\partial{f} }{\partial{p_j}}\biggr),
\end{equation}
where
$f(x,p,t)$ is the probability density of a particle in the phase space at the moment of time $t$;
$m$ is the mass of the particle;
$V(x)$ is the potential function of external forces acting on the particle;  $\gamma=\beta /m$ is the
resistance coefficient of the medium in which the particle moves, per unit of its mass;
$k$ is the Boltzmann constant; $T$ is the temperature of the medium.

Then, instead of equation (\ref{eq_diff1}) for the wave function $\varphi(x, p,t)$, we consider the modified
Kramers equation of the form
\begin{equation}\label{eq_diff2}
\frac{\partial\varphi}{\partial{t}}=A\varphi
+\gamma B{\varphi},
\end{equation}
\begin{equation}\label{def_A2}
\mbox{where }\ \ \ \ \ \ A\varphi =\sum_{j=1}^{n}
\biggl(
\frac{\partial V}{\partial x_j} \frac{\partial\varphi}{\partial p_j}-
 \frac {p_j}{m} \frac{\partial \varphi}{\partial x_j}
\biggr)
-\frac{i}{\hbar}
\biggl(mc^2+V-\sum_{j=1}^{n}\frac{p^2_j}{2m} \biggr)\varphi \ \ \ \ \ \ \
\end{equation}
\begin{equation}\label{def_B2}
\mbox{and }\ \ \ \ \  B{\varphi}=
\sum_{j=1}^{n}\frac{\partial}{\partial{p_j}}\left( \biggl( p_j +i\hbar \frac{\partial}{\partial{x_j}}\biggr){\varphi}
+kTm\frac{\partial{\varphi} }{\partial{p_j}}\right). \ \ \ \ \ \ \ \nonumber
\end{equation}
Equation (\ref{eq_diff2}) is obtained from the Kramers equation (\ref{eq_kramers}) by adding to the right
hand side of the summand of the form
$-{i}/{\hbar} (mc^2+V-{p^2}/(2m) )\varphi$, and the replacement, in the diffusion operator,
of multiplication of the function $\varphi $ by $p_j$
by the action of the operator $(p_j+i\hbar \partial/{\partial x_j})$ on the function $\varphi $.

Adding the summand $-{i}/{\hbar} (mc^2+V-{p^2}/(2m) )\varphi$ is related with the additional physical requirement
that the wave function at the point $(x, p)$ oscillates harmonically with frequency
${1}/{\hbar} (m c^2+V-{p^2}/(2m) )$ in time.

The requirement of harmonic oscillating of the wave function $\varphi $ at the point $(x, p)$ with the
large frequency ${1}/{\hbar} (m c^2+V-{p^2}/(2m) )$, in the case when $ m c^2$ is much greater than
$V,$ leads to the fact that the shift of the wave function with respect to the coordinate $x_j$
with conservation of the proper time at the point $(x, p)$ yields the phase shift in the oscillation of the
function $\varphi$. And the operator of infinitely small shift $\partial/{\partial x_j} $ is changed by the
operator $\partial/{\partial x_j}-ip_j/\hbar$. (For a more detailed explanation, see \cite{ben3}.)
Respectively, if we multiply this operator by $i\hbar$, then we obtain the operator
$p_j+ i\hbar \partial/{\partial x_j}$, used in the modified diffusion operator $B$.

For equation (\ref{eq_diff2}), in \cite{ben4} we obtain results similar to that of the paper \cite{ben3}.
It is shown that also in this case, the process described by equation
(\ref{eq_diff2}), for large $\gamma =\beta /m$
passes several stages.  During the first rapid stage, the wave function goes to a
``stationary'' state of the same form as for equation (\ref{eq_diff2}).
At the second, slow stage,  the wave function evolves in the subspace of ``stationary'' states
subject to the Schrodinger equation. Besides that, it is shown that at the third stage, the
dissipation of the process leads to decoherence of the wave function, and any superposition
of states comes to one of eigenstates of the Hamilton operator.

In the paper \cite{ben4}, it is shown also that if, on the contrary, the medium resistance per unit of mass
of the particle $\gamma =\beta /m$ is small, and in equation (\ref{eq_diff2}) one can neglect the summand
with the factor $\gamma, $
then in the considered model, the density of the probability distribution $\rho =|\varphi |^2$ satisfies
the standard Liouville equation
\begin{equation}\label{liuvill}
\frac{\partial \rho }{\partial{t}}=\sum_{j=1}^{n}
\biggl(
\frac{\partial V}{\partial x_j} \frac{\partial \rho}{\partial p_j}-
 \frac {p_j}{m} \frac{\partial \rho}{\partial x_j}
\biggr),
\end{equation}
as in classical statistical mechanics.

\section{Gauge transformations}
In this section we introduce and discuss the notion of gauge invariance for equation (\ref{eq_diff2}).

According to the approach exposed in \cite{ben4}, the density of probability distribution $\rho(x, p, t)$
of a quantum particle whose state at the moment of time $t$ is given by the wave function $\varphi(x,p,t)$,
is proportional to $ | \varphi |^2=\varphi(x,p,t) \varphi^*(x,p,t).$
This implies that the replacement of a wave function $\varphi  $
by the wave function of the form $\exp (ig/\hbar)\varphi  $,
where $ g=g(x,p,t)$ is an arbitrary real valued function, does not change the density of the probability distribution
$\rho(x, p, t)$.  Such a transform of wave function is usually called a gauge transform.

Let us look how equation (\ref{eq_diff2}) changes under this gauge transform. To this end, let us write out
equation (\ref{eq_diff2}) in a more general form.
Let us write in it, instead of the differentiation operators $\partial /\partial p_j $ of the function $\varphi$,
the operator $D^p_j=\partial /\partial p_j+ i  B_j/\hbar, $   instead of the operators
$\partial /\partial x_j -ip_j/\hbar$, the operator $D^x_j=\partial /\partial x_j +iA_j/\hbar,$
and instead of the operator $\partial /\partial t + iH/\hbar$, where $H=mc^2+p^2/(2m)+V$, let us write the operator
$D^x_0=\partial /\partial t + iA_0/\hbar$, where $A_j, A_0, B_j$  are functions
of $x, p$, and $t$ for $j=1,..., n$. In these notations, equation (\ref{eq_diff2}) will take the form
\begin{equation}\label{cal_diff}
D^x_0\varphi  =\sum_{j=1}^{n}\biggl(\frac{\partial H}{\partial x_j} D^p_j{\varphi}
- \frac{\partial H}{\partial p_j}D^x_j{\varphi}\biggr)+
\gamma \sum_{j=1}^{n}D^p_j\left(i\hbar D^x_j{\varphi}
+kTmD^p_j{\varphi}\right).
\end{equation}

By a gauge transform of equation (\ref{cal_diff}) we call the following transform of the function $\varphi $ and the
potentials $A_j, A_0, B_j$, for $j=1, ..., n$:
\begin{eqnarray}\label{cal_preobr_g}
\varphi &\longmapsto& \varphi'=\exp(-\frac {i}{\hbar} g)\varphi;  \\
A_0&\longmapsto& A'_0= A_0+\frac{\partial g}{\partial t};\nonumber\\
A_j&\longmapsto& A'_j= A_j+\frac{\partial g}{\partial x_j}, \mbox{     where    } j=1, ..., n; \nonumber\\
B_j&\longmapsto& B'_j= B_j+\frac{\partial g}{\partial p_j}, \mbox{     where    } j=1, ..., n.\label{cal_preobr}
\end{eqnarray}

It is not difficult to see that after the substitution (\ref{cal_preobr_g})
into equation (\ref{cal_diff}), replacement (\ref{cal_preobr}), and dividing both parts of the obtained
equality by $\exp(-(i/\hbar) g )$, the form of equation~(\ref{cal_diff}) will not change.

Geometrically, gauge transformation corresponds to transfer to another trivialization of a complex line bundle
over the phase space, in which a form of linear connection is chosen, defining parallel transport
of the vectors of the bundle along trajectories in the phase space.

In the particular case for equation (\ref{eq_diff2}), the potentials read
$$A_0=H(x,p)=E+V;\ \ A_j=-p_j;\ \ B_j=0\ \mbox{for}\ j=1, ..., n.
 $$

Understanding the physical sense of the potentials in the general case for equation (\ref{cal_diff}), requires separate
investigation. For the Dirac equation, potentials of gauge invariance are usually related with the potentials of
electromagnetic field.

\section{The Galileo invariance}
In this section we study the change of equation (\ref{eq_diff2}) under the transfer to a coordinate system
moving uniformly with respect to the initial coordinate system, with the velocity $u$.
The diffusion equation (\ref{eq_kramers}) is not invariant with respect to Galileo transforms under transfer
to new inertial coordinate system moving with constant velocity $u$ with respect to the old one.

The aim of this section is to study invariance of equation (\ref{eq_diff2}) for a free particle ($V=0$)
 with respect to Galileo transforms, with gauge transforms of the wave function.

By definition of Galileo transforms, the new coordinate system is expressed through the old one by the following formulas:
\begin{eqnarray}\label{new_coordinats}
&&t'=t;\ \ \ x'=x-ut;\ \ \ p'=p-mu;\ \ \nonumber\\
 &&E'=\frac{p'^2}{2m}=\frac{(p-mu)^2}{2m}=\frac{p^2}{2m}-pu+\frac{mu^2}{2}=E-pu+\frac{mu^2}{2}.
\end{eqnarray}

Respectively, the old coordinates are expressed through the new ones by the following formulas:
\begin{eqnarray}\label{old_coordinats}
&&t=t';\ \ \ x=x'+ut;\ \ \ p=p'+mu;\ \ \nonumber\\
&&E=\frac{p^2}{2m}=\frac{(p'+mu)^2}{2m}=\frac{p'^2}{2m}+p'u+\frac{mu^2}{2}=E'+p'u+\frac{mu^2}{2}.
\end{eqnarray}

Substituting these expressions into equation (\ref{eq_diff2}), with the use of relations (\ref{def_A2}) and (\ref{def_B2}),
we obtain:
\begin{eqnarray}\label{eq_diff'}
\frac{\partial\varphi}{\partial{t'}}-\sum_{j=1}^{n}\frac{\partial\varphi}{\partial{x'_j}}u_j&=&\sum_{j=1}^{n}
\biggl(
\frac{\partial V}{\partial x'_j} \frac{\partial\varphi}{\partial p'_j}-
 \frac{p'_j+mu_j}{m}
 \biggl(\frac{\partial}{\partial{x'_j}}-i\frac{ p'_j +mu_j}{\hbar }\biggr){\varphi}
\biggr)\nonumber\\
& &-\frac{i}{\hbar}
\left(E'+p'u+\frac{mu^2}{2}+V \right)\varphi \nonumber \\
& &+\sum_{j=1}^{n}\frac{\partial}{\partial{p'_j}}\left( \biggl( p'_j+m u_j +i\hbar \frac{\partial}{\partial{x'_j}}\biggr)
{\varphi}
+kTm\frac{\partial{\varphi} }{\partial{p'_j}}\right), \nonumber
\end{eqnarray}
whence, after simple algebraic transformations, we obtain:
\begin{eqnarray}\label{eq_diff''}
\frac{\partial\varphi}{\partial{t'}}&=&\sum_{j=1}^{n}
\biggl(
\frac{\partial V}{\partial x'_j} \frac{\partial\varphi}{\partial p'_j}-
 \frac{p'_j}{m}
 \biggl(\frac{\partial}{\partial{x'_j}}-i\frac{ p'_j +mu_j}{\hbar }\biggr){\varphi}
\biggr)\nonumber\\
& &-\frac{i}{\hbar}
\left(E'-\frac{mu^2}{2}+V \right)\varphi \nonumber \\
& &+\sum_{j=1}^{n}\frac{\partial}{\partial{p'_j}}\left( \biggl( p'_j+m u_j +i\hbar \frac{\partial}{\partial{x'_j}}\biggr)
{\varphi}
+kTm\frac{\partial{\varphi} }{\partial{p'_j}}\right).\nonumber
\end{eqnarray}

If in the obtained equation one makes the substitution
$\varphi=\exp(({i}/{\hbar})g)\varphi',$ where $g=mux'+mu^2t'/2,$ then (after the gauge transform) we obtain the equation
\begin{eqnarray}\label{eq_diff'''}
\frac{\partial\varphi'}{\partial{t'}}&=&\sum_{j=1}^{n}
\biggl(
\frac{\partial V}{\partial x'_j} \frac{\partial\varphi'}{\partial p'_j}-
 \frac{p'_j}{m}
 \biggl(\frac{\partial}{\partial{x'_j}}-i\frac{ p'_j}{\hbar }\biggr){\varphi'}
\biggr)\nonumber\\
& &-\frac{i}{\hbar}
\left(E'+V \right)\varphi' \nonumber \\
& &+\sum_{j=1}^{n}\frac{\partial}{\partial{p'_j}}\left( \biggl( p'_j +i\hbar \frac{\partial}{\partial{x'_j}}\biggr)
{\varphi'}
+kTm\frac{\partial{\varphi'} }{\partial{p'_j}}\right), \nonumber
\end{eqnarray}
which coincides with equation (\ref{eq_diff2}). Thus, we have proved the Galileo invariance of equation (\ref{eq_diff2}).

\section{Program of further research}
In this section we list directions of further research and sketch approaches to the stated problems.
\newline

{\bf\large5.1. Comparison of the model of scattering of waves with the quantum model}

In order to compare exactness of the model described by equation (\ref{eq_diff2}), with the standard quantum mechanical
model, one should find the situation in which these models give essentially different results.
Such a situation can arise, for example, if one considers the process with the rapidly changing in time potential function
$V(x,t)$. Such a potential can prevent a wave function of equation (\ref{eq_diff2}) from transfer,
during the time of the transition process, to the ``stationary'' one.
As a result, a solution of equation (\ref{eq_diff2})  can differ from a solution of the Schrodinger equation.

In order to check this, consider, for example, the potential function $V=V_0(x)+V_1(x) \cos(\omega t)$ for
$\omega \longrightarrow \infty.$

Mechanical and quantum mechanical systems with such potential were studied in many papers, for example,
\cite{kapitza, landau, cook, grozdanov, gillary, rahav, bandy}.
The physical problem in which such a quantum model arises, is a charged particle in external force field
and in a laser row.

Equation (\ref{eq_diff2}) with this potential reads
\begin{equation}\label{eq_diff3}
\frac{\partial\varphi}{\partial{t}}=A\varphi
+\gamma B{\varphi},
\end{equation}
where
\begin{eqnarray}\label{def_A3}
A\varphi =\sum_{j=1}^{n}
\biggl(
\frac{\partial (V_0+V_1 \cos(\omega t))}{\partial x_j} \frac{\partial\varphi}{\partial p_j}-
 \frac {p_j}{m} \frac{\partial \varphi}{\partial x_j}
\biggr) \nonumber\\
-\frac{i}{\hbar}
\biggl(mc^2+V_0+V_1 \cos(\omega t)-\sum_{j=1}^{n}\frac{p^2_j}{2m} \biggr)\varphi,
\end{eqnarray}
\begin{equation}\label{def_B3}
\mbox{and }\ \ \ \ \  B{\varphi}=
\sum_{j=1}^{n}\frac{\partial}{\partial{p_j}}\left( \biggl( p_j +i\hbar \frac{\partial}{\partial{x_j}}\biggr){\varphi}
+kTm\frac{\partial{\varphi} }{\partial{p_j}}\right). \ \ \ \ \ \ \ \nonumber
\end{equation}

One should study solutions of this equation for large $\omega $ and compare these solutions with solutions of the quantum
system.
\newline

{\bf\large5.2. The study of the scattering process of mixed waves and computation of the time of the transition process
to stationary mixed state of heat equilibrium}

Another problem which one would like to study is the behavior of the process for mixed waves of the form
$\varphi(x, p,t, \xi )$, where
$\xi \in D $ is an additional parameter, and the distribution $\rho (x, p, t)$  in the phase space at the moment
$t$
 for a particle whose state is described by a wave function $\varphi(x, p, t,\xi )$,  is proportional
 to the function $\int_D |\varphi(x, p, t, \xi )|^2d\xi, $
i.~e.
$$\rho (x, p, t)=\frac{\int_D\varphi(x, p,  t,\xi )\varphi^*(x, p,  t,\xi )d\xi}
{\int_{R^{2n}}\int_D\varphi(x, p,  t,\xi )\varphi^*(x, p, t, \xi )d\xi dx dp}.$$
Also here one assumes that the evolution of the wave function in time goes according to
equation (\ref{eq_diff2})
for each fixed $\xi \in D. $

Another equivalent way to describe this process, familiar in quantum mechanics, is to consider
self-adjoint operators $\hat \rho$ on functions on the phase space $R^{2n}$ with the kernel of the operator
of the form
$\hat \rho(x,p;x',p',t)=\int_D\varphi(x, p,  t,\xi )\varphi^*(x', p',  t,\xi )d\xi $. Note that any
positive self-adjoint operator
$\hat \rho$ on the space of functions can be reduced to diagonal form and therefore to the form above. Positive
self-adjoint operators with trace unity are called operators of density of states.
Then the density of probability distribution
$ \rho(x,p ,t)=\hat \rho(x,p;x,p,t)/T\! r \hat \rho,$ where $T\!r \hat \rho=\int_{R^{2n}}\hat \rho(x,p;x,p,t) dx dp$
is the trace of the operator $ \hat \rho$.
The evolution of the operator of density of state $\hat \rho$  in time is given by the equation
$$\frac{\partial \hat \rho }{\partial  t}=
{\cal D}\hat{\rho} +\hat{\rho}{\cal D}^* -\hat \rho T\!r({\cal D}\hat \rho +\hat \rho{\cal D}^*),
$$
where $\cal D$ is the operator expressed by the right hand side of equation (\ref{eq_diff2}), and $\cal D^*$
is the adjoint operator. Expression with the trace $T\!r$ stands in this equation in order to make the trace of
operator of density of state $\hat \rho$ equal to one at each moment of time.

This is a nonlinear equation. One should investigate whether it has a unique stationary state, determine the form
of this stationary state (the state of heat equilibrium), and estimate the time of transition process to this
stationary state.
\newline

{\bf\large5.3. Generalization of the model with account of spin of a particle and the requirement of relativistic
invariance}

Let $M=R^4$ be the Minkowsky space-time, $P=R^3$ be the space of momenta, and $B=M\times P$ be the phase
space-time, on which the Lorentz group naturally acts (if one fixes the stationary mass
$m$ of the particle). The same space has an action of the commutative group of coordinate shifts preserving the
proper time at each point of the phase space,   and of the one-para\-meter group of shifts of proper time
at each point of the phase space. Together these groups define an action of the Poincare group $P$ on the space $B$.

In this new model, we propose to consider the values of the wave function $\varphi $ not in the field of
complex numbers $C$, but in certain Euclidean vector space $F$  over the field of complex numbers, with an action
by unitary linear operators of the group $SU(2,C)$, the two-fold covering of the rotation group
$SO(3)$
of three-dimensional space, acting on the phase space $R^6$. The probability distribution $\rho (x, p, t)$
of position of a particle in the phase space at the moment of time $t$ is again assumed to be
proportional to $|\varphi (x,p, t)|^2.$

The group $SU(2,C)$ is a subgroup in the group $SL(2,C)$, where $SL(2,C)$ is the group of two-dimensional
complex matrices with determinant equal to 1. The group $SL(2,C)$ is the two-fold covering of the
Lorentz group $L$. Thus, we have a commutative diagram of homomorphisms of groups:
$$
\begin{array}{ccccc}
SU(2,C) & \subset & SL(2,C) & \subset & \hat P\\
\downarrow\lefteqn{j} & & \downarrow\lefteqn{j} & & \downarrow\lefteqn{j}\\
SO(3) & \subset & L  & \subset & P,
\end{array}
$$
where $\hat  P$ is the two-fold covering group for the Poincare group.

Further one considers the bundle $pr: F\times B\rightarrow B$ with fiber $F$ over the phase space-time $B$. The
Poincare group $P$ acts on the base $B$. The action of its subgroup $SO(3) \subset L$ on $B$ by rotations
with respect to the coordinate origin lifts to the compatible action of the group
$SU(2, C)$ in the fiber $F$ over the origin point in $B$. Then, this action can be uniquely extended to
an action of the group $\hat P$ on the bundle $F\times B$, compatible with the action of the group $P$ on the
base $B$. The compatibility of the actions of the groups on the bundle means that for any $g\in \hat P$, the following
diagram is commutative:
$$\begin{array}{ccc}
F\times B & \stackrel{g}{\longrightarrow}& F\times B \\
\downarrow\lefteqn{pr}&& \downarrow\lefteqn{pr}\\
B&\stackrel{j(g)}{\longrightarrow}& B.
\end{array}
$$
Here also, if $g\in SU(2,C)\subset \hat P$, then the diagram
$$\begin{array}{ccc}
F\times \bar 0&\subset&F\times B \\
\downarrow\lefteqn{g}&&\downarrow\lefteqn{g}\\
F\times \bar 0&\subset&F\times B.
\end{array}
$$
is commutative.

The uniqueness of the lift of the action of the Poincare group from $B$ to the action of the group
$\hat P$ on the bundle $F\times B$ is understood up to a choice of trivialization of this bundle.

If $a, b \in B$ are two points of the base (the phase space-time), then one uniquely defines an element $h_{a,b}\in P$
of the Poincare group, of the parallel transport of the coordinate system from the point $a$ to the point $b$.
The action of the element $h_{a, b}$ lifts uniquely to the action of an element $\hat h_{a, b}\in \hat P$
on the bundle $F\times B$. This action transfers elements of the fiber $F$ over $a$ to elements of the fiber
over $b$. Let us call this action by the parallel transport of elements of the fiber along the vector $\vec {ab}$.
Further, this definition allows us to define the parallel transport in the bundle $F\times B$ along any curve
in the base.

In the considered model, the wave function $\varphi $ at the moment of time $t $  is given by a function on the
phase space of the form $\varphi : R^6 \rightarrow F$. The evolution of the wave function in time is defined
by the condition that it is simultaneously in several motions:

1) The vector $\varphi(x,p) \in F$  is parallel transported along the trajectory in the phase space;
the trajectory is defined by a random Brownian process according to certain diffusion equation, for example,
the Kramers equation.

2) The vector $\varphi $  at each point $(x,p)$, in the coordinate system related to this point, rotates with the
constant angular velocity $\omega =m c^2/\hbar$ in the fiber $F$  over this point in the proper time related to this
point; the direction of the rotation axis
$J_{x, p}\in su(2,C)$, in the stationary (laboratory) coordinate system, transforms from one point to another
in the same way as the direction of the angular momentum.

The value of the wave function at the point $(x, p)$  at the moment $(t+\triangle t)$ is defined by the mean value
of the vectors $\varphi $
over all trajectories ending at the point $(x, p)$ of the phase space at the moment $(t+\triangle t)$.

One should construct the differential equation corresponding to this model, and study it.\newline

{\bf\large5.4. Scattering of waves on the phase space and interaction with electromagnetic field}

This problem is related to introducing interaction with electromagnetic field into the model.
Such introducing could be made by analogy to its introducing into the Dirac equation.
As it was shown in equation (\ref{cal_diff}), on this way potentials arise depending also on the momentum,
in contrast with the vector potential of the electromagnetic field which depends only on coordinates and time.
Determining the sense of vector potentials depending on momenta, also requires a separate investigation.

{\bf Acknowledgements:}  
I am grateful to  professor A.~V.~Stoyanovsky, who translated this paper to English.

\end{document}